\documentclass[aip,pre,reprint,superscriptaddress]{revtex4-1}
\usepackage{amssymb}
\usepackage{amsmath}
\usepackage{caption}
\usepackage{graphicx}
\usepackage{epstopdf}
\newcommand{\appropto}{\mathrel{\vcenter{
  \offinterlineskip\halign{\hfil$##$\cr
    \propto\cr\noalign{\kern2pt}\sim\cr\noalign{\kern-2pt}}}}}

\allowdisplaybreaks
\begin{document}
 \title{Convenient analytical formula for cluster mean diameter and diameter dispersion after nucleation burst}
 \author{M. Tacu}
 \affiliation{ \'Ecole Normale Sup\'erieure Paris-Saclay, 94230, Cachan, France}
 \affiliation{Princeton Plasma Physics Laboratory, Princeton University, Princeton, New Jersey, 08543}
 \author{A. Khrabry *}
 \affiliation{Princeton  Plasma Physics Laboratory, Princeton University, Princeton, New Jersey, 08543}
 \author{I.D. Kaganovich}
 \affiliation{Princeton Plasma Physics Laboratory, Princeton University, Princeton, New Jersey, 08543}

 \date{\today}
 \begin{abstract}
We propose a new method of estimating the mean diameter and dispersion of clusters formed in a cooling gas, right after the nucleation stage. Using a moment model developed by Friedlander [S.K. Friedlander, Ann. N.Y. Acad. Sci. 354 (1983)], we derive an analytic relationship for both cluster diameter and diameter dispersion as a function of two of the characteristic times of the system - the cooling time and primary constituents collision time. These formulas can be used to predict diameter and dispersion variation with process parameters such as the initial monomer pressure or cooling rate. It is also possible to use them as an input to the coagulation stage, without the need to compute complex cluster generation during the nucleation burst. We compared our results with a nodal code and got excellent agreement.
 \end{abstract}
 \maketitle

\section{Introduction}

A considerable effort was made to understand nanoparticle formation and evolution in a cooling gas. For applications \cite{1,2} using these nanosized particles, it is important to understand which parameters determine resulting mean particle size and its dispersion. While we know since pioneering Girshick's works \cite{3,4} that the final diameter of nanoparticles formed in a cooling gas is affected by the gas pressure and the cooling rate, a quantitative relationship has been missing. The present paper aims at obtaining analytical formulas which allow a direct computation of the first two moments of the particle size distribution after the nucleation stage.

Cluster growth in a cooling gas can be divided into two stages\cite{4,5}: the nucleation and growth stage.  As the initially stable (non supersaturated) gas cools down, the vapor saturation pressure decreases below the gas pressure and the cluster formation process begins. With further temperature drop, this process intensifies as the saturation pressure decreases precipitously with the temperature, faster than gas pressure. The supersaturated gas returns to equilibrium via the nucleation burst - a phase of rapid cluster/droplet nucleation, when the barrier to their formation can be overcome at sufficiently low temperature. Metal clusters form at temperatures of the order of few thousands Kelvin. This high temperature is typically achieved in plasma arc or torch or laser ablated plasmas \cite{6,7,8}.
 
The time elapsed before the clusters are generated in a nucleation burst and corresponding value of the supersaturation degree are crucial parameters describing the process of cluster formation and growth. Once substantial amount of clusters is formed and most of the monomers are consumed, the growth stage begins, in which the clusters grow by merging with each other via coagulation \cite{9}.

A good description of the  classical nucleation theory (CNT) is given in Refs.  \cite{10,11,12} and Ref.\cite{13} reviews experimental studies. Note that CNT is a phenomenological theory and have a number of limitations. Therefore making quantitative predictions using this theory or its recent modifications is still  subject to debate as pointed out in the references above and Ref.\cite{14} . Main limitation of CNT arises because of the application of bulk macroscopic properties to very small clusters, among which the surface tension that cannot be defined for clusters containing only few atoms.\cite{12}

We will recall the main physical ideas of CNT and use an appropriate version of CNT to give simple, yet accurate analytical estimates of the mean cluster diameter and its dispersion  after the nucleation burst. These estimates give scaling law as a function of gas pressure and cooling rate applicable for a wide range of gas and material parameters.  One could then use these estimates for the mean cluster diameter and its dispersion as an initial input for a coagulation model. That way there is no need to simulate evaporation condensation process during nucleation, which involves taking into account all the cluster sizes and makes straightforward simulation of cluster size evolution a very cumbersome numerical problem.  
\hfill \break 

\textit{Brief discussion of the classical nucleation theory} 

\hfill \break 
During the gas cooling when gas becomes supersaturated and monomers associate to form clusters, small clusters form first and later grow by absorbing more and more monomers. However, formation of small clusters is energetically unfavorable. There is an energy barrier \cite{15} $\Delta \Phi = \Phi - \Phi_0$, where $\Phi$ is the thermodynamic potential of the system $\{ $vapour, liquid droplets$\}$ and  $\Phi_0$ the potential of the system before the liquid droplet formation. The change in the potential due to cluster (liquid droplet) formation is :  
\begin{equation}
\begin{gathered}
\Delta \Phi =\mu_ln_l+\mu_gn_g+4\pi r^2\gamma-\mu_g(n_g+n_l)\\
= -(\mu_g-\mu_l)n_l+4\pi r^2\gamma\\
= -\frac{\mu_g-\mu_l}{N_A}N+\epsilon_sN^{2/3}
\end{gathered}
\end{equation}
Here, $N$ is the number of monomers (primary constituents) in the cluster, $n_l$ and $n_g$ are the amount of liquid and gas, in moles, in the final state (the total amount of matter $n_g+n_l$ is conserved) and $N_A$ is the Avogadro number. The molar liquid chemical potential is denoted by $\mu_l$ and the molar gas chemical potential by $\mu_g$. 

In the last equation, the first term in the right-hand side (RHS) corresponds to the binding energy of atoms within the liquid volume. The second term in the RHS corresponds to the surface energy (it is proportional to the cluster surface area or $N^{2/3}$), which, in fact represents an effect of the binding energy reduction for atoms at the cluster's surface. These atoms do not bond as strong as other atoms deep in the liquid. That's why this term has an opposite sign to the first term. Here, $\epsilon_s$ is the specific surface energy of the cluster which can be deduced from the surface tension coefficient $\gamma$ , by 
\begin{equation}
\epsilon_s=4\pi r_W^2 \gamma, 
\end{equation}
where $r_W$ is the Wigner-Seitz radius defined so that  $4\pi r_W^3\rho/3m_a=1$, where $\rho$ is the mass density of liquid and $m_a$ is the mass of a monomer.

For an ideal gas and incompressible liquid, the chemical potential difference can be written\cite{15}  by introducing Boltzmann's constant $k$ :
\begin{equation}
    \mu_l-\mu_g = kN_A T ln(S),
\end{equation}
where the supersaturation degree $S$ is defined using $n_1$ - the monomer's number density in the gas and $n_{sat}$ - the number density corresponding to the saturation conditions : 
\begin{equation}
S=\frac{P}{P_{sat}} = \frac{n_1}{n_{sat}}. \label{S}
\end{equation}
Here, $P$ is the actual pressure and $P_{sat}=n_{sat}kT$ is the saturation pressure at gas-liquid equilibrium over a flat surface.
We define the initial time ($t=0$), so that $S=1$  at that moment and we set $n_{sat}(t=0)=n_1(t=0) \equiv n_0$.
The saturation particle density is then given by Clausius-Clapeyron law :
\begin{equation}
n_{sat}(T)T = {{n_0T_0}}e^{\frac{e_a}{k}{\left(\frac{1}{T_0}-\frac{1}{T}\right)}\label{n_sat}},
\end{equation}
where $T$ is the actual temperature and $e_a$ is the vaporization energy per atom for a flat surface. 

Because we only consider here short nucleation stage, we can assume a constant gas cooling rate $\dot{T}_0$ and linear decrease of temperature with time:   
\begin{equation}
T(t)=T_0-\dot{T}_0\times t. \label{T(t)}
\end{equation}

\begin{figure}
\includegraphics[scale=0.6]{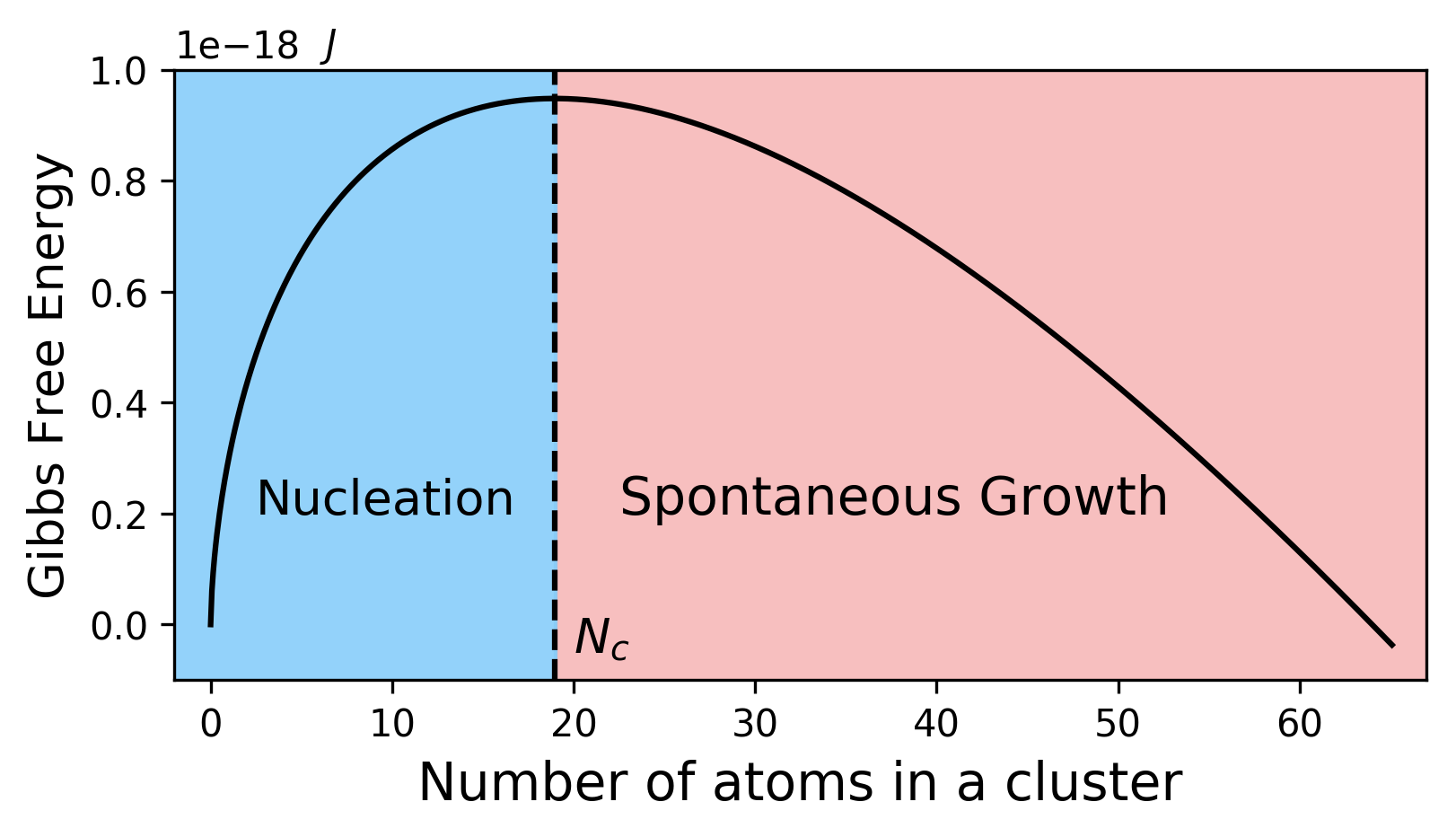}
\caption{$G(N)$ - the Gibbs free energy of formation of a cluster containing $N$  atoms as a function of $N$} \label{Fig:G_n}
\end{figure}

As expected, at first, the supersaturation degree increases with time, because the monomer gas density remain nearly constant (no monomer consumed yet), but saturated gas density $n_{sat}$ decreases with temperature.   According to this picture, we can approximately write (using Eq.(5) : $\ln(n_0/n_{sat}) \approx \frac{e_a\dot{T_0}}{kT_0^2}t$), which allows us to express the supersaturation (see Fig.3) :
\begin{equation}
    S(t) \approx \begin{cases}
    \exp\left(\frac{e_a\dot{T_0}}{kT_0^2}t\right), & t<t_0, \\
    1, & t>t_0.
    \end{cases}
\end{equation}
Here, $t_0$ is time when the nucleation burst occurs (at the supersaturation's maximum) or simply nucleation time when nearly all monomers are quickly consumed into clusters.

The determination of this time, which is a critical parameter of the nucleation stage, will allow us to express the mean cluster diameter and its dispersion which is the main goal of this paper. 

We  modeled cluster formation numerically using the so-called Nodal General Differential Equation (NGDE\cite{16}) and also using Friedlander's model detailed in the next section. The two methods agree well as shown in Fig.2 and Fig.3.  The exponential behaviour of the supersaturation with rapid decrease to unity at nucleation was already observed, e.g.,  in Refs. \cite{3,4}.
\hfill \break 

\textit{The homogeneous nucleation rate} 

\hfill \break
The thermodynamic potential $\Phi$ where the pressure, temperature and number of particles are used, is the Gibbs free energy\cite{17} $G(N)$ , the typical profile of which is shown in Fig.\ref{Fig:G_n}. The function is non-monotonic, for small clusters the free surface energy (the second term in the RHS of Eq.(1)) dominates over the binding energy (the first term in the RHS of Eq.(1)) and the Gibbs energy $G(N)$ is growing with $N$. At some value of $N$, commonly referred as the critical number $N_c$ ,
\begin{equation}
N_c= \left(\frac{2\epsilon_s}{3 kT lnS}\right)^3\label{N_c},
\end{equation}
the function reaches its maximum and then monotonically decreases.
Corresponding cluster size is called critical cluster diameter, $d_{cl}^*=2r_W N_c^{1/3}$,
\begin{equation}
d_{cl}^*(S,T)=r_W\frac{4\epsilon_s}{3 kT lnS}\label{d_c}.
\end{equation}

\begin{figure}
\includegraphics[scale=0.67]{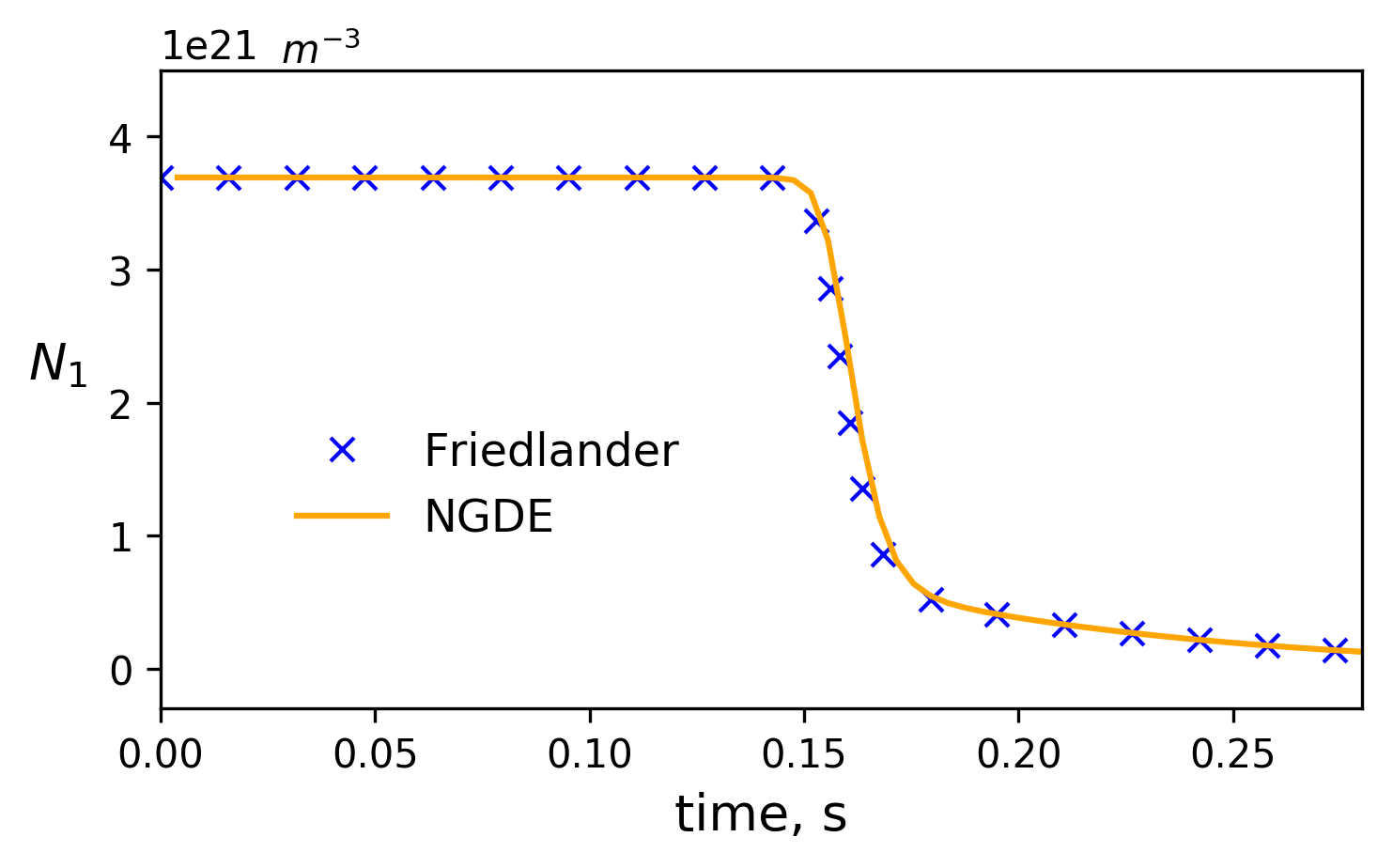}
\caption{Monomer density $n_1$ as a function of time for aluminum with the same parameters as in Fig.3. NGDE code used 41 nodes (b) Mean diameter of clusters calculated with Friedlander's model (crosses) and  NGDE code (orange line).}
\end{figure}
In  other words, for small clusters with a number of  atoms less than $N_c$, growth is energetically unfavorable (attachment of each next atom to the cluster results in the Gibbs energy increase), but for larger clusters with a number of atoms larger than $N_c$, the cluster growth is energetically favorable. Hence, once a cluster has reached a critical size, it will spontaneously grow by consuming the gas monomers, absorbing them on its surface. But to reach the critical size, a cluster needs to overcome (by thermal fluctuations) the energy barrier:
\begin{equation}
\begin{gathered}
\frac{G(n_c)}{kT} = 4\theta^3/27\ln(S)^2,\\
\theta=\frac{e_s}{kT}.
 \end{gathered}
\end{equation}
Therefore we are mostly interested in critical cluster production, since they will continue growing (the critical diameter decreasing before nucleation burst), but their production is a slow process, especially when $S$ is not very big. The derivation of their production rate is the main object of the CNT.

As pointed out in Bakhtar's paper \cite{11}, CNT was developed by Volmer, Becker and Doring, Zeldovich and others. A lot of modifications were proposed, accounting for Gibbs free energy corrections, non steady distribution for clusters smaller than the critical size, etc. Unfortunately, despite a century of research, there is still opportunity for a definitive nucleation theory to emerge\cite{10}. One of the main points of content\cite{18} of the CNT is the capillarity approximation which extends the bulk thermodynamic properties to nano-scale clusters, which in turn leads to errors in estimating the free energy of small clusters. The comparison with molecular dynamics simulations are given in Ref.\cite{19}

\begin{figure}
\includegraphics[scale=0.67]{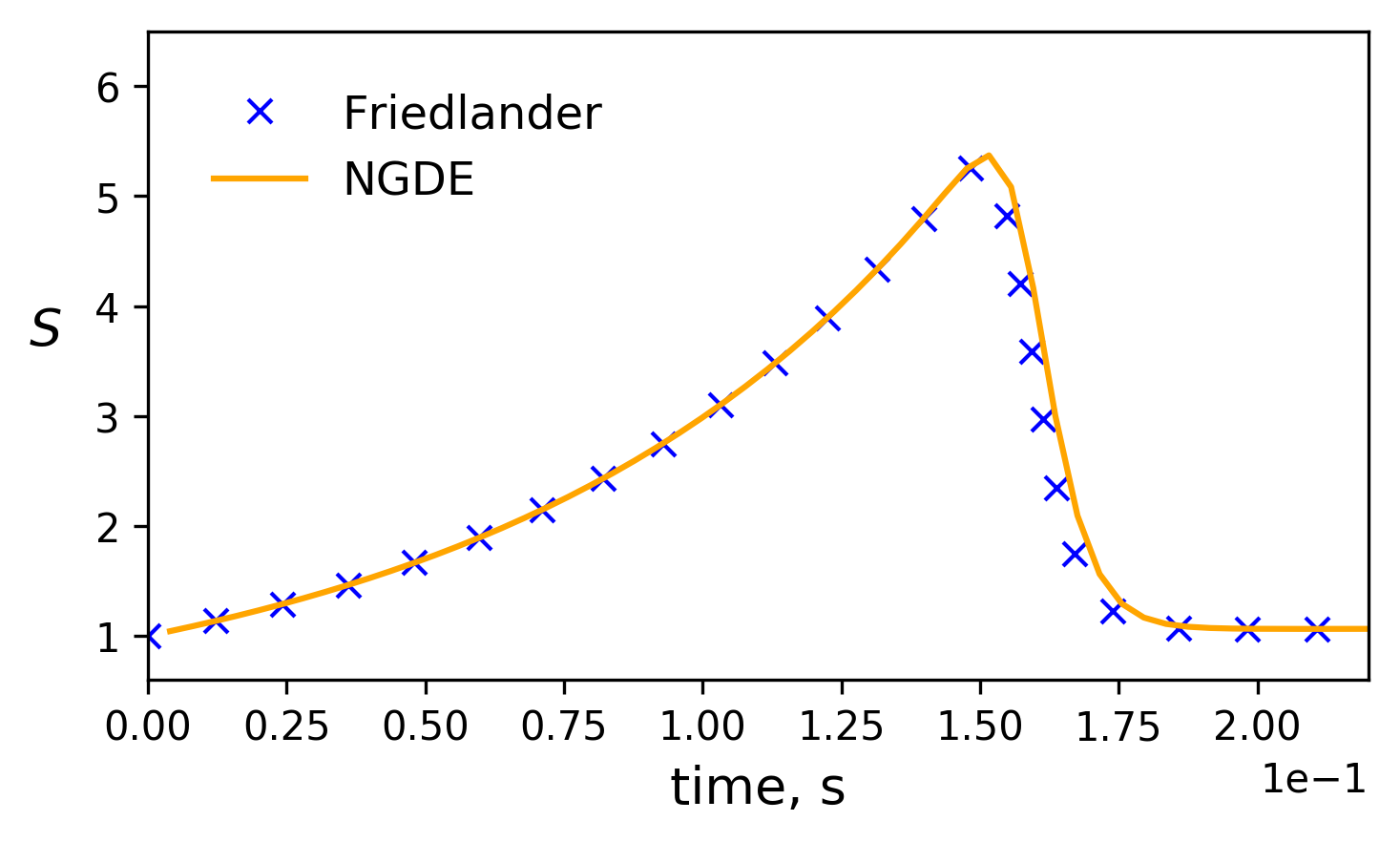}
\caption{The supersaturation $S$ as a function of time for aluminum and cooling rate $\dot{T}_0=1000$K/s, at  pressure $P_0=60Pa $,  $T_0=1773$K and $\gamma$=0.948N/m calculated with Friedlander's model (crosses) and  NGDE code (orange line).}
\end{figure}
There were several attempts to correct this problem\cite{14}, but the most consistent correction (consistent does not necessarily mean experimentally accurate) is Girshick's one \cite{14,20} which gives following expression for the rate of cluster production of critical size $N_c$, or nucleation rate:
\begin{equation}
\begin{gathered}
J=n_1 n_{sat}v_1 \sqrt{\frac{2\gamma}{\pi m_a}}e^{\theta-4\theta^3/27\ln(S)^2},\\
\approx J(t_0)e^{2b_0(t-t_0)/t_0^3},\\
b_0=\frac{8e_s^3T_0}{27e_a^2k\dot{T_0}^2}.
 \end{gathered}
\end{equation}
The nucleation rate and its approximation as an exponential function are given in Appendix 1 (see also Fig.10).
Here, $m_a$ is monomer mass. We also introduce the average volume in liquid per monomer
\begin{equation}
v_1=4\pi r_W^3/3, 
\end{equation}
and average surface in liquid per monomer
\begin{equation}
a_1=4\pi r_W^2. 
\end{equation}
The nucleation rate is a very strong function of $S$ and has only substantial values for large $S$. When the supersaturation is low ($S \approx 1$), the condensation does not start immediately because the nucleation rate is yet negligible. After substantial amount of clusters have been produced, monomers rapidly condense on cluster surfaces, and the number density of monomers rapidly decreases. As a result, the gas reaches equilibrium with the liquid in clusters and the supersaturation degree $S$ drops to unity. The further cluster growth occurs via cluster agglomeration and Ostwald ripening processes.

Here we use the NGDE code that solves the General Differential Equation \cite{9} (GDE) accounting for condensation on and evaporation from clusters and also for their agglomeration. NGDE codes typically use a logarithmic discretization of cluster volume space, and therefore the NGDE solution is subject to numerical diffusion due to rough discretization of the cluster size distribution. 

General GDE solvers are computationally intensive. Different type of codes such as the Kinetic Monte Carlo\cite{22} code fail to simultaneously simulate both the nucleation and the cluster growth, because these processes are on rather different time scales, the nucleation event being very short comparing to coagulation time. It is possible to make assumptions on the particle size distribution, such as supposing it lognormal for example \cite{21}, but this is not the case for all the systems \cite{9}.

\section{Friedlander's Model}\label{s2}

In order to derive an analytical expression for $t_0$ we use a moment model for the cluster size distribution as derived by S.K. Friedlander. 

The Friedlander's moment model is a system of equations for the first three moments of cluster size distribution function $f(d_{cl})$: 
\begin{equation}
n_{cl} \equiv \int_{d_{cl}^*}^{\infty}f(d_{cl})d(d_{cl}),
\end{equation}
\begin{equation}
M_1 \equiv \int_{d_{cl}^*}^{\infty}  d_{cl} f(d_{cl})d(d_{cl}),
\end{equation}
\begin{equation}
A \equiv \int_{d_{cl}^*}^{\infty} \pi d_{cl}^2 f(d_{cl})d(d_{cl}).
\end{equation}
Here, $n_{cl}$, is the number density of clusters above the critical diameter, $A=<\pi d_{cl}^2>n_{cl}$ relates to average surface area of clusters above the critical diameter and $M_1=<d_{cl}>n_{cl}$ relates to the average diameter of clusters above the critical diameter.

During short nucleation stage the agglomeration process can be neglected, because it happens on a much longer time scale than the evaporation/condensation process. The cluster diameter in this case is simply given by the assumption of an uniform growth of clusters above critical size :
\begin{equation}
\frac{d(d_{cl})}{dt} = 2 (n_1-n_{sat}) v_1 v_{th}. \label{evaluation d}
\end{equation}
Here, $v_{th}=\sqrt{\frac{kT}{2\pi m_a}}$ is the thermal velocity of monomers.
Substituting Eq.(\ref{evaluation d}) into definitions of momenta above gives \cite{1} :
\begin{equation}
\frac{dn_{cl}}{dt}=J  \label{n_{cl}},
\end{equation}
\begin{equation}
\frac{dM_1}{dt}=2r_W N_c^{1/3}J +2 v_1v_{th} (n_1-n_{sat})n_{cl},  \label{M1}
\end{equation}
\begin{equation}
\frac{dA}{dt}=a_1 N_c^{2/3}J +4\pi v_1v_{th}(n_1-n_{sat})M_1, \label{A}
\end{equation}
\begin{equation}
\frac{dn_1}{dt}=-N_c J - v_{th} (n_1-n_{sat}) A \label{N1}.
\end{equation}
Here, Friedlander neglected the terms proportional to  $f(d_{cl}^{*}) d(d_{cl}^{*})/dt$, because at the beginning of the nucleation $d_{cl}^{*}$ is nearly constant and at the end $f(d_{cl}^{*})$ is small, therefore the contribution of the product is small. Also, at the nucleation burst $d(d_{cl}^{*})/dt$ vanish exactly because of $S$ reaching its maximum.

Eq. (18) describes the evolution of the clusters density, $n_{cl}$ (again only clusters above critical size are considered). Clusters of critical size form at a rate $J$ which is determined by Eq. (11). Moreover, all the clusters formed above critical size stay above critical size because they only grow and never reduce in size, whereas critical size reduces with time as S increases, see Fig.3. Agglomeration of clusters is not considered in the model, which is a valid simplification for nucleation stage.

Eq. (\ref{M1}) describes size evolution of the clusters' average diameter. $M_1$ is the first moment of the clusters size distribution. The first term in the right-hand side (RHS) accounts for formation of new clusters of critical size. The second term in the RHS describes the clusters growth via atoms deposition on cluster surfaces. In this model we neglect the dependence of the deposition flux on a cluster size and assume the flux for a flat surface, that is flux is proportional to $(n_1-n_{sat})$.
It is convenient that clusters diameter growth rate depends only on the deposition flux and does not depend on a cluster size. This is equivalent to ignoring the effect of surface curvature on saturation pressure (the Kelvin effect). In our case is not necessary to take it into account, since it is small for clusters bigger than the critical size wich we consider here. NGDE simulations confirm this assumption. Net deposition flux is derived as the difference between evaporation and deposition fluxes: $v_1v_{th} (n_1-n_{sat})$. 

Eq. (\ref{A}) describes evolution of the clusters average area (second moment of the size distribution). $A$ is clusters total surface area within a gas volume unit. As in equations (\ref{M1}) and (\ref{N1}), the first term in the RHS stands for formation of new critical size clusters, the second term in the RHS accounts for the surface deposition. 

Eq. (\ref{N1}) is a monomer balance which describes decay of the monomers density due to formation of new clusters and the gas condensation on the surface of existing clusters.

We performed simulations with both the Friedlander's model and the NGDE solver for an example of Aluminum vapor cooling with $\dot{T}_0=1000$ K/s and $T_0=1773$ K. At the initial moment saturated gas is considered ($S=1$). Fig. 2 and Fig. 3 shows the comparison between two models. There is a very good agreement between the Friedlander's model and the full general differential equation, showing that agglomeration of clusters (which is neglected in the Friedlander's model but accounted in NGDE) does not play a significant role during the nucleation stage. 

We also verified the negligible role that the carrier gas plays in our case. We used Wedekind\cite{23} work to change the nucleation rate and observed negligible change in the final diameter although there were some change in the nucleation rate. We conclude that the thermalization with the carrier gas is sufficiently rapid\cite{24} to keep nucleation under isothermal conditions.

\begin{center}
\begin{tabular}{|c|c|c|c|c|}
  \hline
  Material & $kT_0/e_a$&$\theta_0$&$e_s/e_a$&$\tau_{collision}$\\
  \hline
  Al & $0.08$ &5&0.4&$1.5 \times 10^{-8}$s\\
  Au & $0.07$ &9&0.7&$4.1 \times 10^{-8}$s\\
  Ag & $0.08$ &9&0.7&$2.8 \times 10^{-8}$s\\
  Cu & $0.08$ &10&0.8&$2.9 \times 10^{-8}$s\\
  B  & $0.06$ &4&0.2&$1.9 \times 10^{-8}$s\\
  \hline
\end{tabular} 
\captionof{table}{Values of the main parameters for different materials at atmospheric pressure and corresponding Clapeyron temperature}
\end{center}

\begin{figure}
\centering
\includegraphics[scale=0.66]{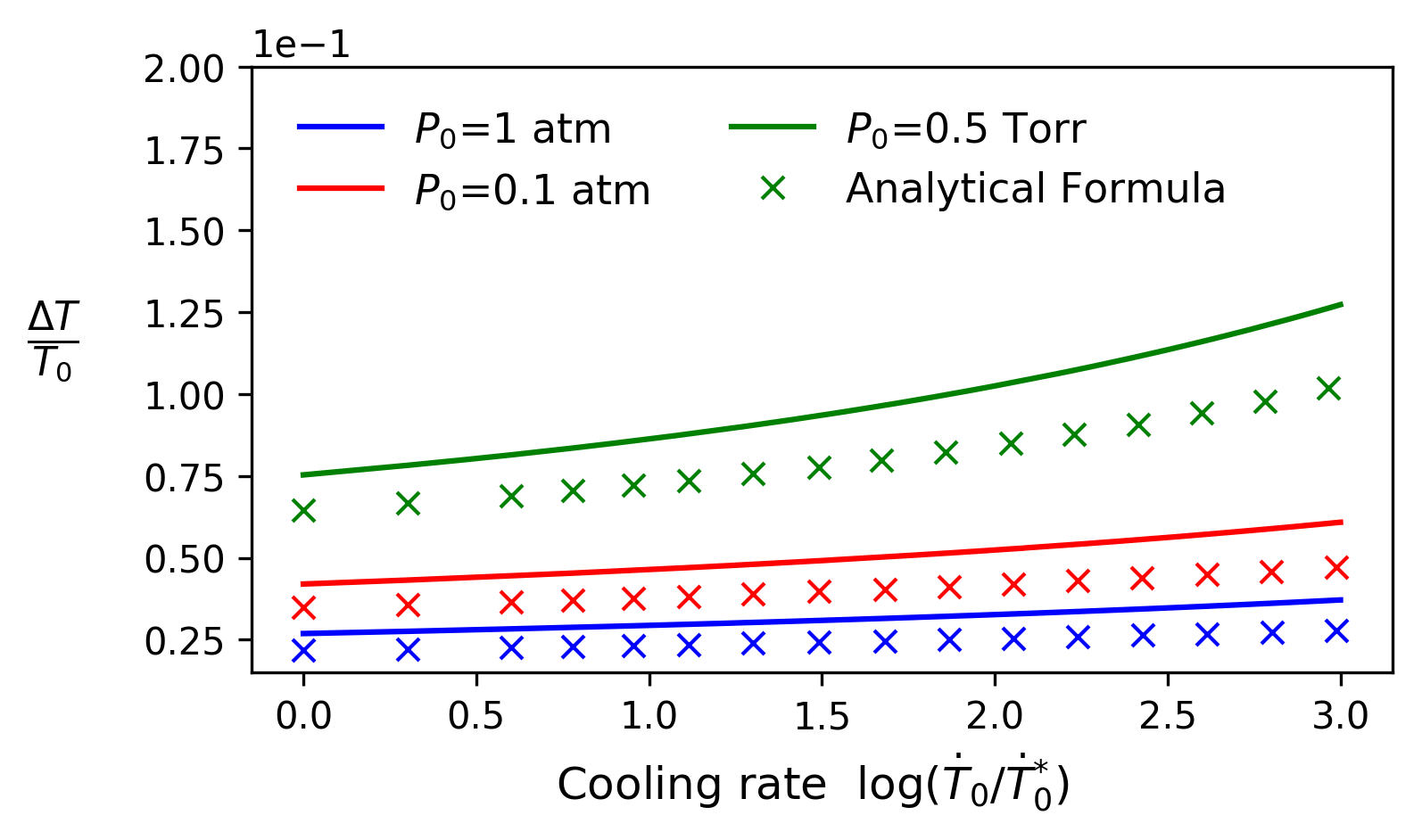}
\caption{$\frac{\Delta T}{T_0}$ as a function of the cooling rate from Friedlander (lines) and from the analytical formula (crosses) in a log scale at different $P_0$ and with $\dot{T_0}^{*}=$ 1000K/s.}
\end{figure}
\section{Analytic expression for the nucleation time}\label{s3}
In this section we give an analytic expression to the crucial parameter that describes the nucleation stage - the nucleation time $t_0$. Because of the sharpness of the nucleation event\cite{4}, this time will also give us an estimate of the time at which the transition process between nucleation and coagulation starts.

The time $t_0$ being defined as the time at which the supersaturation $S$ reaches its maximum, we derive in the Appendix 1 an equation of evolution of $S$ from Friedlander's model : 
\begin{equation}
\frac{dS}{dt} - \frac{\dot{T_0} e_a}{kT_0^2}S + (S-1)v_{th}A=0
\end{equation}
At the beginning of the cooling, $A$ - the total area of clusters above the critical size, is small since the critical size is infinite ($S$ close to 1) and clusters can not durably form via monomer attachment. The term containing $A$ can thus be neglected which results in an exponential growth of $S$.

\begin{figure}
\centering
\includegraphics[scale=0.67]{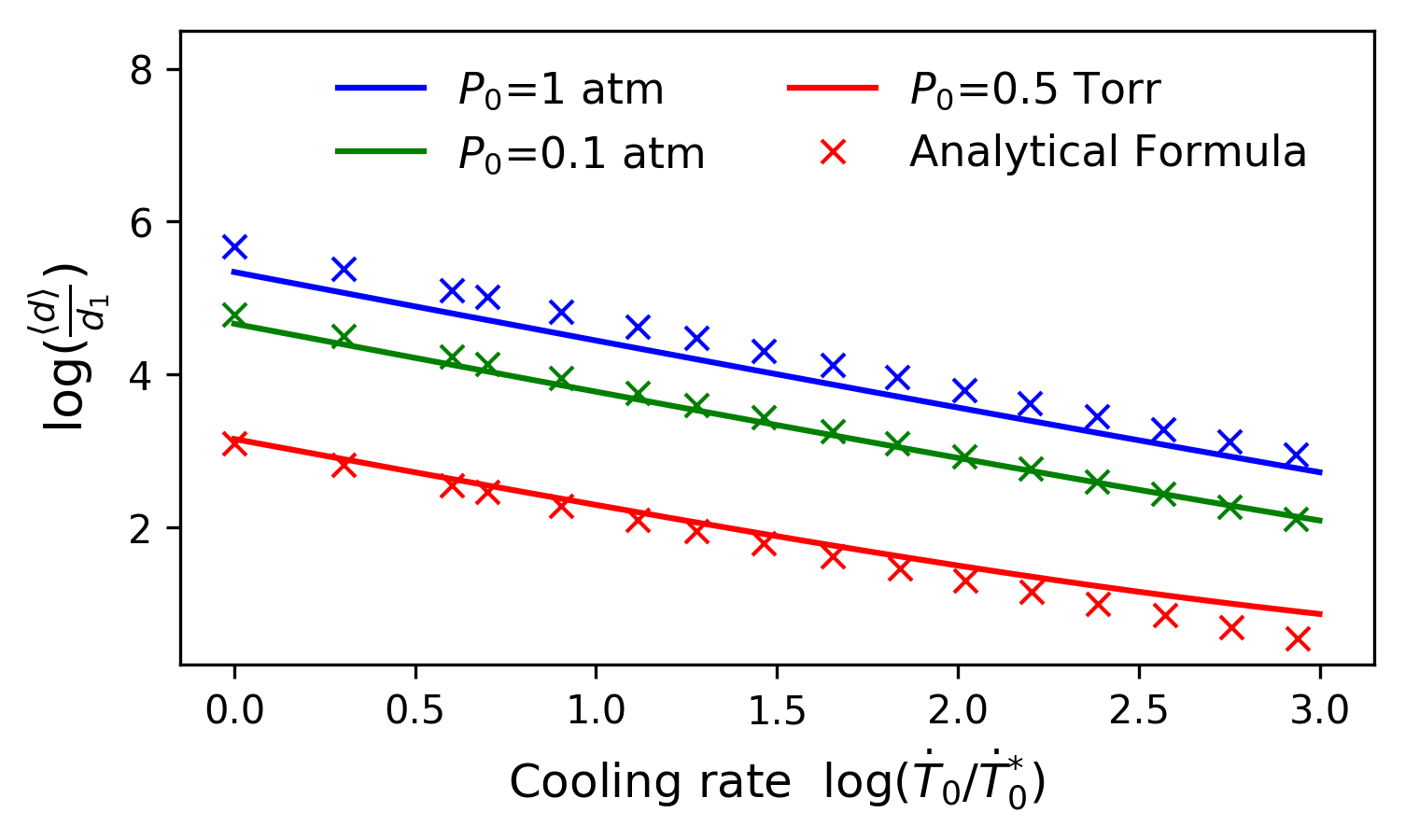}
\caption{Comparison in log-log scale between the analytic diameter (crosses) and the diameter from Friedlander's model (lines) as a function of the cooling rate with $\dot{T_0}^{*}=1000$ K/s and $d_1=2r_W$.}
\end{figure}

When $A$ becomes sufficiently big, because of it growing as the triple integral of $J$ (thus exponentially), we can neglect the middle term. It leads us to an equation on $S-1$, which rapidly decrease to 0.

Using Eq.(22), we found an analytical formula in good agreement with the simulations as shown in Fig.4. It involves the $W$ Lambert's function (scipy.special.lambertw in Python) and is given in non dimensional terms by :
\begin{equation}
\begin{gathered}
\frac{\Delta T}{T_0} =  \frac{kT_0}{e_a}\left( \frac{\theta_0}{3}\right)^{3/2}\left[ W\left(\frac{\tau_{cooling}}{\tau_{collision}}\right)\right]^{-1/2},\\
\tau_{cooling} = \frac{T_0}{\dot{T_0}}, \\
\tau_{collision} = \frac{1}{v_0V_1^{2/3}n_0}
\end{gathered}
\end{equation}
where $v_0 = v_{th}(t=0)$ and $\theta_0 = \theta(t=0)$.

Here $\Delta T$ is the temperature difference between the initial time when the saturation pressure equals to the gas pressure and the nucleation time $t_0$ :
\begin{equation}
\Delta T = T_0-T(t_0) = \dot{T}_0 t_0
\end{equation}

We observe that the dimensionless quantity $\Delta T/T_0$ is given by a slowly varying Lambert function, as a ratio between the two of the characteristic times of the system, the cooling time and the monomer collision time.

One could use the slow variations of $\Delta T/T_0$ with respect to monomer gas pressure or cooling rate to rapidly estimate the nucleation time as a function of the cooling rate and temperature. The variations of $t_0$ can also be inferred, the nucleation time is nearly inversely proportional to the cooling rate, as already observed in Ref. \cite{3}

The formula was tested for Aluminium in Fig.5, where it is compared with predictions from Friedlander's model. It also shows good agreement for other materials given in Table 1.

\begin{figure}
\centering
\includegraphics[scale=0.66]{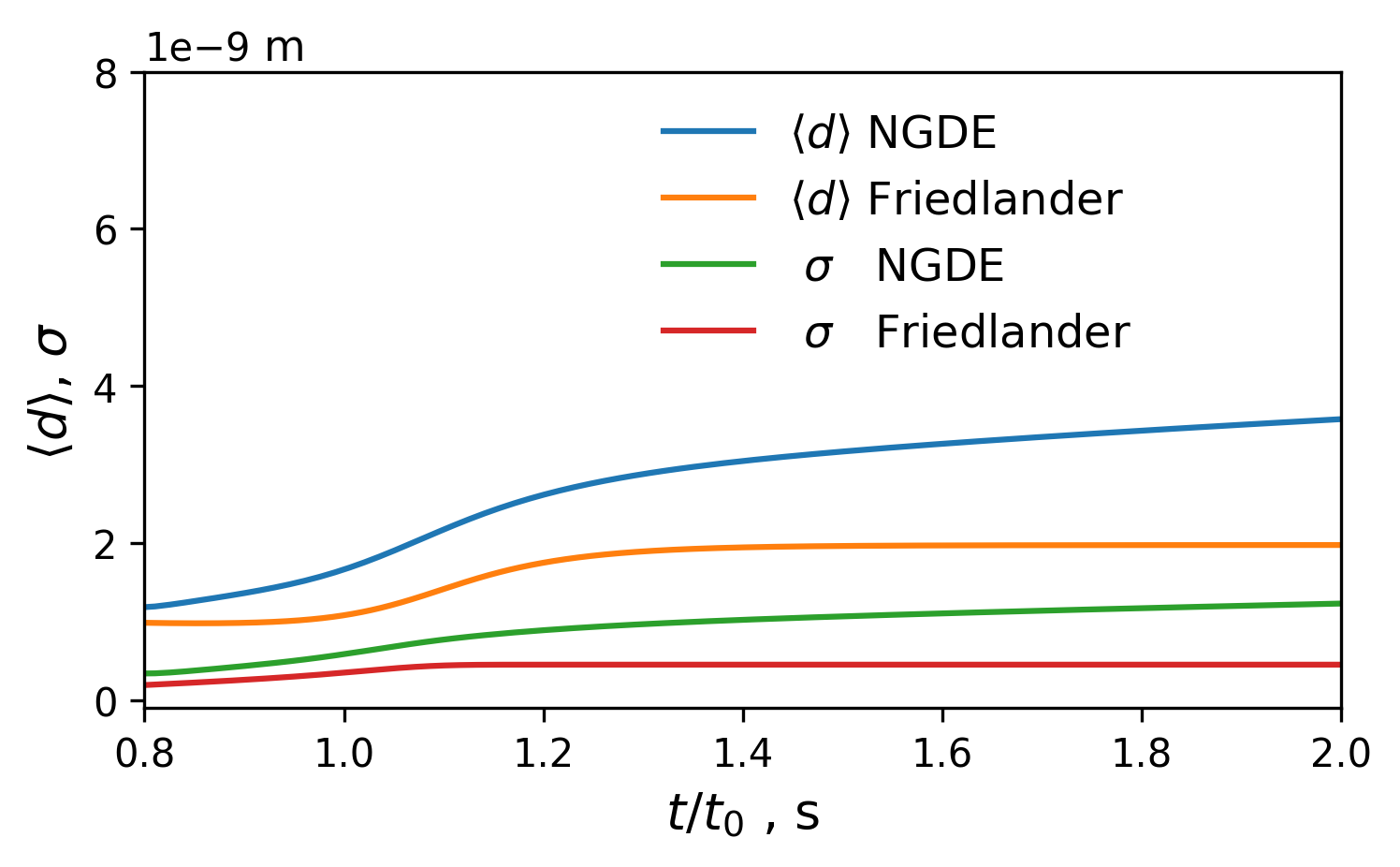}
\caption{The first two moments of the particle size distribution $f(d)$ for Aluminium at $T_0=1773$ K and $\dot{T_0}=10^6$ K/s}
\end{figure}

\begin{figure}
\centering
\includegraphics[scale=0.66]{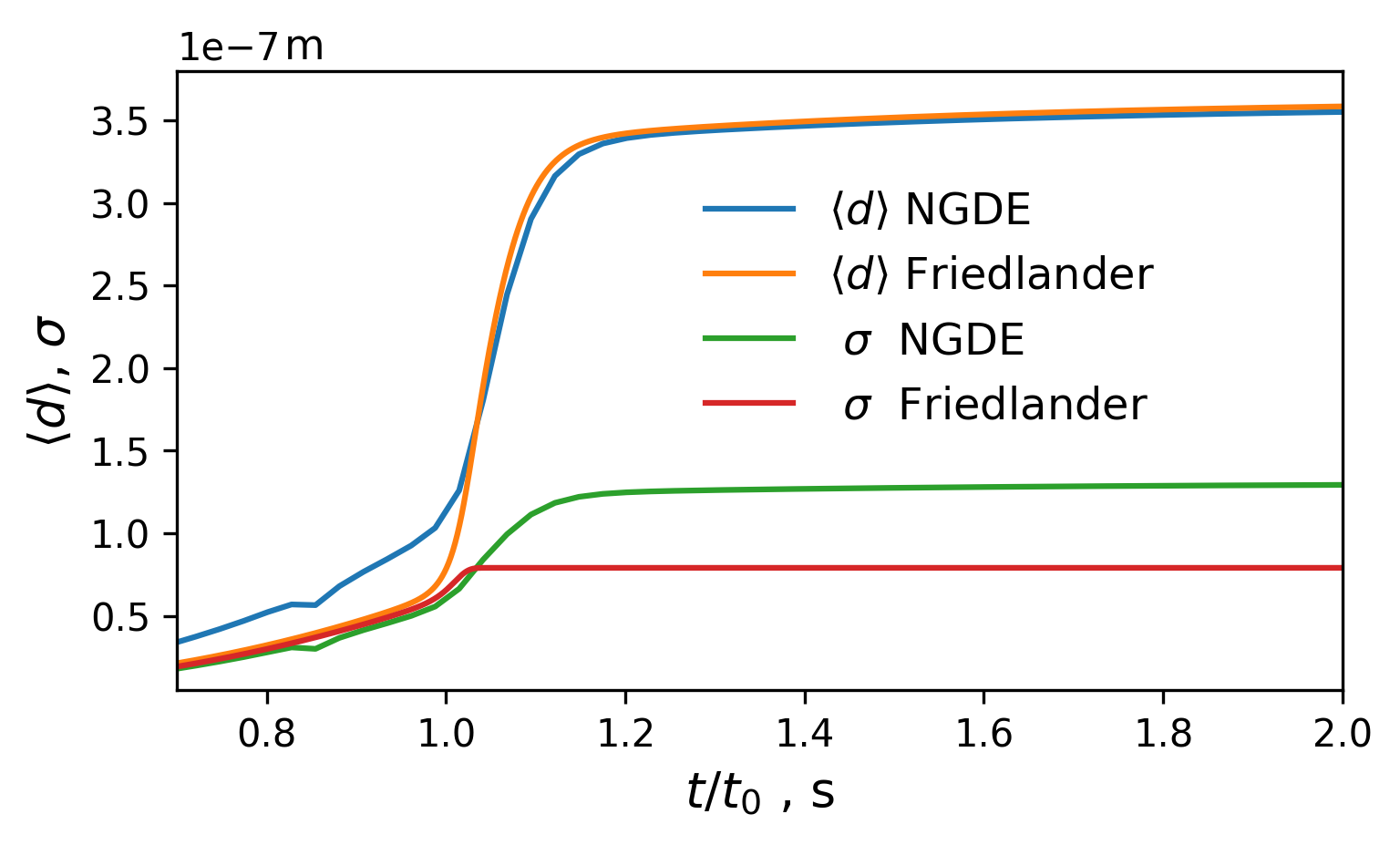}
\caption{The first two moments of the particle size distribution $f(d)$ for Aluminium at $T_0=1773$ K and $\dot{T_0}=10^3$ K/s}
\end{figure}

\section{Diameter and Dispersion}
In this section we express the mean particle diameter $\langle d\rangle$ and dispersion  $\sigma$ at the end of the nucleation stage. We define the diameter as $\langle d\rangle = \frac{M_1}{n_{cl}}$, and the dispersion as   \cite{1} :
\begin{equation}
\sigma^2=\frac{A}{\pi n_{cl}} - \left(\frac{M_1}{n_{cl}}\right)^2
\end{equation}
We expect the dispersion to grow before the nucleation burst takes place and then to reach an asymptote because of the uniform growth for particles bigger than the critical size.

However, as seen in Fig.6 and Fig.7 from NGDE simulations, which unlike Friedlander's model do take into account coagulation, the dispersion is affected by coagulation that happens during and after the nucleation stage.

The coagulation process will depend on the total number of clusters squared ( in the case where we consider a collision between all of the particles ). This coagulation term could be approached roughly by a term proportional to $n_{cl}^2$.

From Eq.(27), $n_{cl}^2$ is bigger for smaller diameter. This explains the more visible coagulation in Fig.6 for $\dot{T_0}=10^6$ K/s, than in Fig.7 for $\dot{T_0}=10^3$ K/s, since in the first case the mean diameter at nucleation is smaller (see Ref.\cite{4} for the influence of the cooling rate on diameter at nucleation), meaning that there will be more clusters in the collision stage. It is also apparent on the particle distribution plots in Fig.8 and Fig.9.

If we look at moment's derivatives obtained from Eq.(18)-(21) :
\begin{equation}
\begin{gathered}
\frac{d \langle d\rangle}{dt}=\frac{J}{N}(d_{cr}-\langle d \rangle) + 2V_1v_{th}n_{sat}(S-1),\\
\frac{d \sigma^2}{dt} = \frac{J}{N^2}\left(N d_{cr}^2-2M_1 d_{cr} -\frac{A}{\pi}+2\frac{M_1^2}{N} \right)
\end{gathered}
\end{equation}
we recognize a nucleation term in the two equations, and an attachment term in the diameter derivative.

When the nucleation has finished, $J$ drops very rapidly to 0, so that only the mean diameter continue to grow because of a monomer deposition on the clusters. It eventually reaches an asymptote, when excess of the monomers from the gas phase has condensed on the clusters and $S$ drops to 1 after the nucleation has finished.

As expected in Friedlander's model, the dispersion $\sigma$ will reach an asymptote immediately after the nucleation. This depedence in $J$ of the derivative of $\sigma$ gives a low dispersion to Friedlander's model.

It is thus possible to compute $\langle d \rangle$ by neglecting the dispersion and using the total number of nucleated clusters $n_{\infty}=n_{cl}(t=\infty)$. At the end, almost all the monomers are attached to clusters so that $n_0/n_{\infty}$ represent the average number of monomers in a cluster. 

From that, it is straightforward to deduce (conservation of matter) :
\begin{equation}
\langle d \rangle = \left(\frac{6V_1n_0}{\pi n_{\infty}}\right)^{1/3}
\end{equation}
We thus only need to compute $n_{\infty} = \int_{0}^{\infty}J(t)dt$  which we know, in the absence of agglomeration to be close to  $\int_{0}^{t_0}J(t)dt$ since almost all the particles come from the nucleation before the nucleation burst. We deduce from slow variations of the $W$ function in Appendix 2, from Eq. (11) :
\begin{equation}
\begin{gathered}
\langle d \rangle= \frac{r_W e_s}{3e_a} \frac{\tau_{cooling}/\tau_{collision}}{W(\tau_{cooling}/\tau_{collision})} \appropto \frac{n_0}{\dot{T_0}},\\
\sigma = \frac{e_a}{e_s}\frac{\Delta T}{T_0}\langle d \rangle
\end{gathered}
\end{equation}
For the materials presented in Table I, $\frac{\Delta T}{T_0} \approx 0.1 $ and $e_a \approx e_s$, which gives us a dispersion an order of magnitude lower compared to the mean diameter. However, even if small, it is not zero. It is the main reason we chose the Friedlander's model over more used ones, as for example Nemchinski monodisperse model \cite{25} or Panda model \cite{5}, accounting for both nucleation and coagulation. With this approach we can compute analytically the dispersion and compare it with full codes such as NGDE.

These formulas give a quantitative explanation to a result already observed by Girshick \cite{3,4}, and can be used either to roughly estimate the final particle size and their dispersion or as an input for a coagulation model, thus without having to compute the evaporation/condensation process.

\begin{figure}
\centering
\includegraphics[scale=0.66]{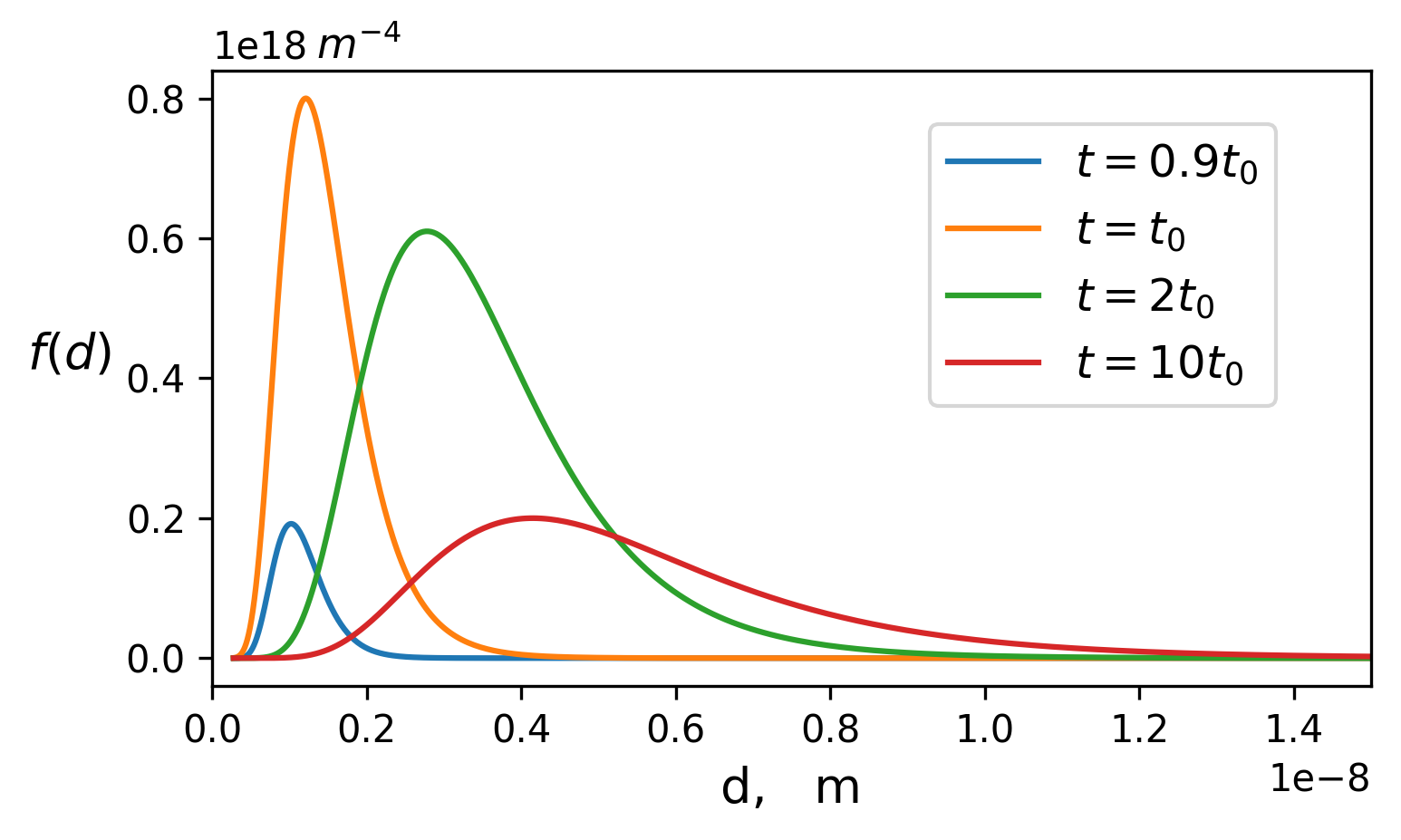}
\caption{Particle size distribution, interpolated as a lognormal function from 41 points (nodes) in NGDE. Data for Aluminium at $T_0=1773$ K, $\dot{T_0}=10^6$ K/s and different times, from $T(t_0)$ up to solidification temperature.}
\end{figure}

\begin{figure}
\centering
\includegraphics[scale=0.66]{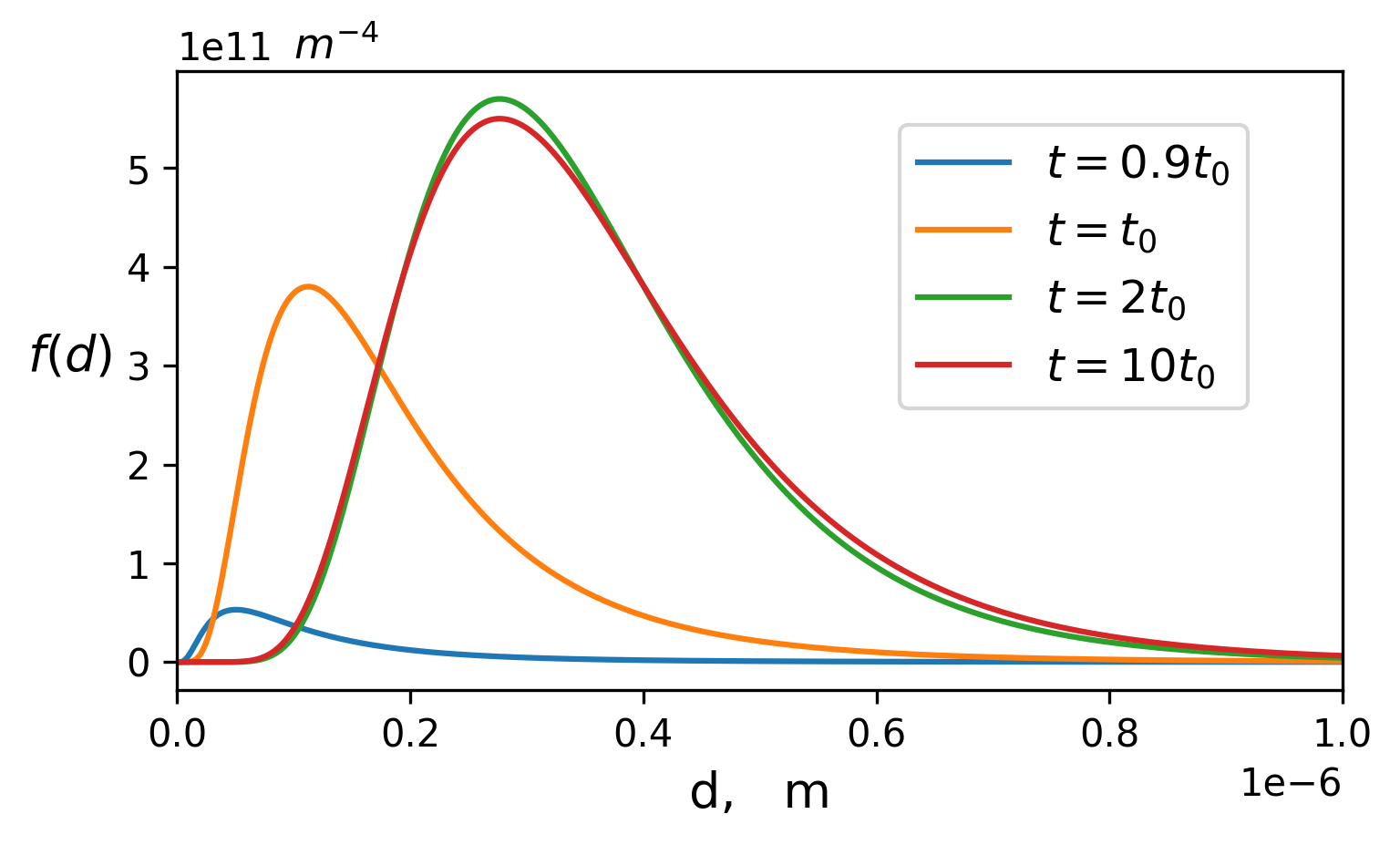}
\caption{Particle size distribution, interpolated as a lognormal function from 41 points (nodes) in NGDE. Data for Aluminium at $T_0=1773$ K, $\dot{T_0}=10^3$ K/s and different times, from $T(t_0)$ up to solidification temperature.}
\end{figure}

\section{Conclusion}

In this study we derived an analytical formula for both mean diameter and its dispersion after the nucleation stage, using a Friedlander's momentum model and matter conservation. We find that the cluster size and its dispersion are proportional to the gas pressure (either carrier gas pressure or monomer pressure, since they are linked via the initial equilibrium condition $S=1$) and inversely proportional to the cooling rate. We also express the nucleation time as a function of the cooling time and the time of Brownian collision between gas particles. We successfully compared our results with a nodal code (NGDE).

\acknowledgements
This work was supported by ENS Paris Saclay and the US Department of Energy.
The authors would also like to thank Edward Startsev, Roscoe B. White, Yevgeni Raitses, and Steven Girshick for very helpful discussions.

*Current affiliation: Lawrence Livermore National Laboratory (LLNL). LLNL is operated by Lawrence Livermore National Security, LLC, for the U.S. Department of Energy, National Nuclear Security Administration under Contract DE-AC52-07NA27344

\section{Appendix 1}
In this section, we obtain an analytical formula for the nucleation time $t_0$ defined as $\frac{dS}{dt}(t=t_0)=0$ using Friendlander's moment equations Eq.(18-21). We first obtain a simplified version of these equations by neglecting the nucleation terms with $J$ in Eq.(19-21).
Let's show that for reasonable cooling rates (typically $\dot{T_0} < 10^6$ K/s) we can neglect the nucleation terms with respect to the deposition terms (containing $n_1-n_{sat}$). Since $n_{cl}=\int_0^{\infty}J(t)dt$ and $J$ is varying over a small time $\delta t$, we can write $n_{cl} \approx J \delta t$, then near the nucleation burst ($t=t_0$) in Eq.(18) :

\begin{equation}
N_cJ \ll v_{th}(n_1-n_{sat})A 
\approx v_{th}n_0a_1n_{cl} \approx v_{th}n_0a_1J \delta t\\
\end{equation}

Neglecting $N_cJ$ with respect to $v_{th}(n_1-n_{sat})A $ is thus equivalent to showing that $N_c \ll v_{th}n_0a_1 \delta t$. Here $v_{th}n_0a_1 \delta t$ is the number of particles attached to a cluster during the nucleation burst (during $\delta t$). For low cooling rates this number is bigger than the critical number at the nucleation burst (where $N_c$ reaches its minimum).

Actually, during the nucleation burst a lot of clusters of critical size are formed and grow essentially from monomer attachment, so the number of monomers that can attach to a particular cluster during this time, or $v_{th}n_0a_1 \delta t$ should be much bigger than $N_c$ - the number of monomers in a nucleated cluster. Numerical simulations confirm this idea. If we compare the ratio between $v_{th}n_0a_1 \delta t$ and $N_c$, we find for aluminium at $T_0=1773K$ : $v_{th}n_0a_1 \delta t / N_c \approx 10$ for $\dot{T_0} = 10^6$ K/s , and  $v_{th}n_0a_1 \delta t / N_c \approx 1000$ when $\dot{T_0} = 10^3$ K/s, as expected.

By neglecting nucleation terms, we simplify Friedlander's model as follows : 

\begin{equation}
\begin{gathered}
\frac{dn_1}{dt}= -v_{th} (n_1-n_{sat})A \\
\frac{dA}{dt}=4\pi V_1v_{th}(n_1-n_{sat})M_1\\
\frac{dM_1}{dt}=2V_1v_{th} (n_1-n_{sat})n_{cl} \\
\frac{dn_{cl}}{dt}=n_1 n_{sat}V_1 \sqrt{\frac{2\gamma}{\pi m_a}}e^{\theta-4\theta^3/27\ln(S)^2}.\\
\end{gathered}
\end{equation}

Let's then assume that $n_1(t)=n_0$ before $t_0$ and see how can we simplify the Friedlander's model.

While replacing $n_1$ by $n_0$, we refer to Fig.3(a) to observe that the monomer's density is almost constant throughout the nucleation process, before the nucleation burst. From a mathematical point of view it is due to our initial conditions in Eq.(19-21) since we set the derivatives of $n_1$ as being 0 at $t=0$ up to the fourth order. Physically, we understand the slow variation of $n_1$ as a consequence of the high energy barrier that the small clusters need to overcome to grow and thus consume monomers.

We replace for simplicity $v_{th}$ by $v_0=v_{th}(t=0)$, since $v_{th}$ is a slowly varying function of $T$, and introduce a dimensionless time by :

\begin{equation}
u=\frac{\dot{T_0}t}{T_0}
\end{equation}

Now with $n_1n_{sat} \approx n_0^2$ and $\theta(t=0)=\theta_0$ we can express $J$ as : 

\begin{equation}
\begin{gathered}
J(u) = n_0^2V_1 \sqrt{\frac{2\gamma}{\pi m_a}}e^{\theta_0}e^{-g(u)}\\
g(u) = \frac{b}{u^2(1-u)} \approx \frac{b}{u^2} \\
b=\frac{4\theta_0^3}{27(e_a/kT_0)^2}
\end{gathered}
\end{equation}

The nucleation rate is now only a function of $u$, which allows us to transform Eq.(32) into a directly integrable system of equations. To do so, let's recall that from Eq.(5) :
$\frac{1}{n_{sat}} \frac{dn_{sat}}{dt} = \frac{\dot{T_0}}{T}\left(1-\frac{e_a}{kT}\right) \approx -\frac{e_a\dot{T_0}}{kT_0^2} $.
This will allow us to link $dS/du$ and $dn_1/du$ to obtain (we derive $S=n_1/n_{sat}$) : 

\begin{equation}
\begin{gathered}
\frac{dS}{du} -\frac{e_a}{kT_0}S + (S-1)\frac{T_0}{\dot{T_0}}v_0A=0\\
\frac{dA}{du}=\frac{T_0}{\dot{T_0}}4\pi V_1v_0(n_0-n_{sat})M_1\\
\frac{dM_1}{du}=\frac{T_0}{\dot{T_0}}2V_1v_0 (n_0-n_{sat})n_{cl}\\
\frac{dn_{cl}}{du}=\frac{T_0}{\dot{T_0}}n_0^2V_1 \sqrt{\frac{2\gamma}{\pi m_a}}e^{e_s/kT}e^{-g(u)}.\\
\end{gathered}
\end{equation}

Even if Eq.(33) is a simplified version of Eq.(30), it is still impossible to integrate it analytically (because of $e^{-g(u)}$). In order to integrate the system, we will develop the nucleation rate near $u_0=\frac{\dot{T_0}t_0}{T_0}$ as an exponential function. To do so, we write $g$ near $u_0$ as $g(u) \approx g(u_0)+(u-u_0)g'(u_0)$ and with $g'(u_0) \approx -2b/u_0^3$, we get:

\begin{equation}
J=J(u_0)e^{-b(u-u_0)g'(u_0)} \approx J(u_0)e^{2b(u-u_0)/u_0^3}
\end{equation}

Now we can explicitly find $N,M_1,A$ and $n_1$ by direct integration of Eq.(33).  We see from Fig.10, the results of our approximations of $J$ and the errors that we get by integrating one time $J$ to obtain $N$. We will however perform this integration to find the nucleation time $t_0$ since the sharp increase of $J$ ensures a small error in the nucleation time, as we will see further.

Now we replace $n_0-n_{sat}(t)$ by $n_0-n_{sat}(t_0) \approx n_0$ following the idea that the coefficients have a slower variation in time than the moments. Replacing the coefficients by their value at $t_0$ allows us to analytically integrate (Eq.33). First it transforms into the system :  

\begin{equation}
\begin{gathered}
\frac{dS}{du} -\frac{e_a}{kT_0}S + \frac{v_0T_0}{\dot{T_0}}A(S-1)=0\\
\frac{d^3A}{du^3}=ae^{2b(u-u_0)/u_0^3}.\\
a= \left(\frac{T_0}{\dot{T_0}} \right)^3 8\pi V_1^2v_0^2n_0^2 J(u_0)
\end{gathered}
\end{equation}

Then, by neglecting the terms in $e^{-2b/u_0^2}$ while integrating $A$ :

\begin{equation}
A(u) = a\left(\frac{u_0^3}{2b}\right)^3e^{2b(u-u_0)/u_0^3}
\end{equation}
From (Eq.35)  with $S_{max}$ being the maximum supersaturation degree $S(u_0)$ and with $\frac{dS}{du}(u_0) = 0$, we get :

\begin{equation}
A(u_0) = \frac{e_a \dot{T}_0}{kT_0^2v_0}\frac{S_{max}}{S_{max}-1} \approx \frac{e_a \dot{T}_0}{kT_0^2v_0}
\end{equation}

Using both Eq.(36) and (Eq.37) we obtain an equation on $u_0$ : 
\begin{equation}
u_0^9J(u_0) = \frac{b^3e_a \dot{T}_0^4}{\pi kT_0^5v_0^3V_1^2n_0^2}
\end{equation}
We could find the nucleation time numerically from the last equation, but we notice that if we approach $u_0^9$ with $(u_0^2)^{9/2} = (b/g(u_0))^{9/2}$ (from Eq.(32)), we can get an explicit formula using the $W$ Lambert's function.

\begin{figure}
\centering
\includegraphics[scale=0.7]{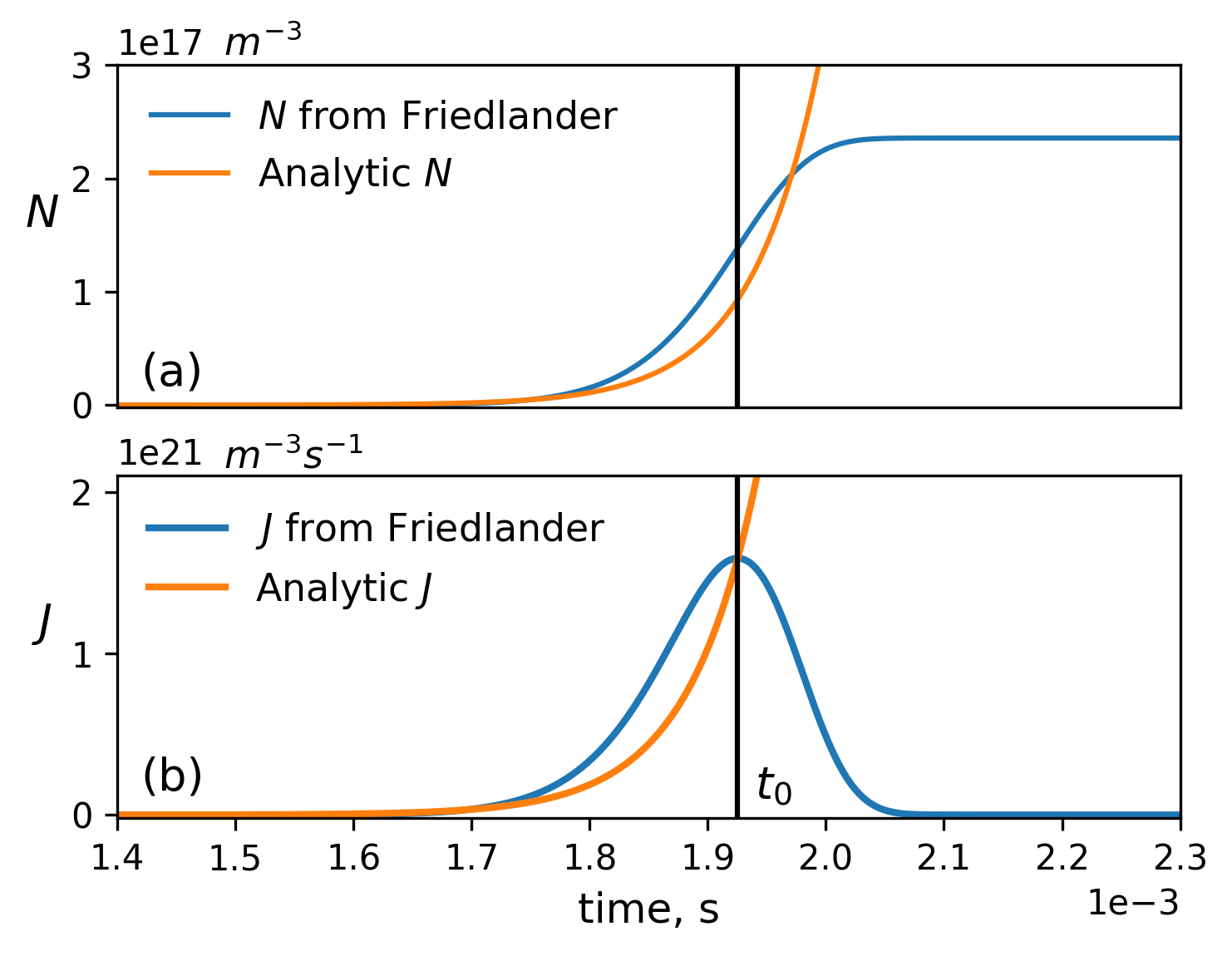}
\caption{Here $\dot{T_0}=10^5$ K/s and $T_0=1773$K (a) The total concentration of clusters $N$.
(b) $J$ from Friedlander and approximated with an exponential}
\end{figure}

\begin{equation}
\left(g(u_0)\right)^{9/2}e^{g(u_0)} = \frac{V_1^{1/3}b^{3/2}\sqrt{\frac{2\pi\gamma}{m_a}}kT_0e^{\theta_0}}{e_av_0} \left(\frac{\tau_{cooling}}{\tau_{collision}}\right)^4 
\end{equation}
By replacing $b$ with its value from Eq.(32) and raising the equation at the power $1/4$ we get : 

\begin{equation}
g(u_0)^{9/8}e^{g(u_0)/4} = \alpha \theta_0^{1/4}e^{\theta_0/4} \frac{e_s\tau_{cooling}}{e_a\tau_{collision}}
\end{equation}

Where $\alpha$ is a numerical factor of $ \alpha = (4\pi/3)^{1/12}(4/27)^{3/8} \pi^{1/8} $ 

By assuming that $g(u_0)^{9/8} \approx g(u_0)$  we get with $4/\alpha \approx 6$ and by neglecting $e_s\alpha\theta_0^{1/4}e^{\theta_0/4}/e_a$ since W is slowly varying :

\begin{equation}
g(u_0) = 4W\left( \frac{e_s\tau_{cooling}}{e_a\tau_{collision}} \right) 
\end{equation}

Finally with the definition of $g$ from Eq.(32) :

\begin{equation}
u_0 = \frac{kT_0}{e_a}\sqrt{\frac{\theta_0^3}{27W(e_s\tau_{cooling}/e_a\tau_{collision})}}
\end{equation}

\section{Appendix 2}
In this section we derive an expression for the mean diameter $\langle d \rangle$ and of the dispersion $\sigma$ after the nucleation burst, when the monomers have condensed on the clusters and S has become equal to 1. This corresponds to the asymptotic values of $\frac{M_1}{n_{cl}}$ and of $\sqrt{A/ \pi n_{cl} - (M_1/n_{cl} )^2}$ from Friedlander's model  Eq.(18-21). We want to integrate $J$ on the interval [0,$u_0$] since the main contribution to the total number of clusters in the absence of coagulation comes from the cluster nucleation, so that, using Eq.(34) for J :

\begin{equation}
\begin{gathered}
n_{\infty} \approx n_{u_0} = \frac{T_0}{\dot{T_0}} \int_0^{u_0}  J(u_0)e^{2b(u-u_0)/u_0^3} du \\
= \frac{T_0 u_0^3}{2b\dot{T_0}} J(u_0) \left(1-e^{-2b/u_0^3}\right)\\
\approx  \frac{T_0 u_0^3}{2b\dot{T_0}} J(u_0)
\end{gathered}
\end{equation}

In Section IV we showed that Friedlander's model should have a low dispersion since the dispersion $\sigma$ stops growing after the nucleation burst. In the limit of zero dispersion, average diameters of the clusters can be expressed via their average volume as follows :

\begin{equation}
\langle d \rangle = \left(\frac{6V_1n_0}{\pi n_{\infty}}\right)^{1/3} \approx  \left(\frac{12V_1n_0b\dot{T_0}}{\pi T_0 u_0^3 J(u_0)}\right)^{1/3}
\end{equation}
Using Eq.(40) we obtain :

\begin{equation}
\begin{gathered}
\langle d \rangle = \left(\frac{3^7}{4}\right)^{1/3}\frac{kT_0V_1 e_a v_0 u_0^2 }{e_s^2} \frac{n_0T_0}{\dot{T_0}} \\
\approx \frac{8kT_0V_1 e_a v_0 u_0^2 }{e_s^2} \frac{n_0T_0}{\dot{T_0}} 
\end{gathered}
\end{equation}
With Eq.(42) we obtain the final expression of the mean diameter at the end of the nucleation stage : 

\begin{equation}
\langle d \rangle= \frac{2r_W}{3} \frac{e_s\tau_{cooling}/e_a\tau_{collision}}{W(e_s\tau_{cooling}/e_a\tau_{collision})}
\end{equation}

Let's now derive the mean dispersion, that according with the section IV will not change after nucleation, so we can express the final dispersion as being at the moment of nucleation burst : 
With Eq.(35) which we integrate two times and neglect the exponential terms, we obtain (the expressions are evaluated in $u_0$ at the nucleation burst) :

\begin{equation}
\begin{gathered}
\sigma^2  =\frac{A}{\pi n_{cl}} - \left(\frac{M_1}{n_{cl}} \right)^2\\
=2V_1^2v_0^2n_0^2 \frac{T_0^2u_0^6}{\dot{T_0}^2b^2} - V_1^2v_0^2n_0^2 \frac{T_0^2u_0^6}{\dot{T_0}^2b^2} \\
 = \left(\frac{27e_a^2 V_1v_0u_0^3 P_0T_0}{4e_s^3\dot{T_0}} \right)^2
\end{gathered}
\end{equation}

Combining Eq.(42) and Eq.(47) yields finally :
\begin{equation}
\begin{gathered}
\frac{\sigma}{\langle d \rangle} =  \frac{27e_a}{32e_s}u_0
\approx \frac{e_a}{e_s}u_0
\end{gathered}
\end{equation}
We can see on Fig.7 how $\sigma \approx \langle d \rangle$ at $t_0$ and how different they are at the end. We can recover this behaviour from the equations above.

\end{document}